# Arrays of optical vortices formed by "fork" holograms

**A. Ya. Bekshaev, A. S. Bekshaev, K. A. Mohammed**

*I.I. Mechnikov National University, Dvorianska 2, Odessa 65082, Ukraine*

*bekshaev@onu.edu.ua*

**Abstract.**
Singular light beams with optical vortices (OV) are often generated by means of thin binary gratings with groove bifurcation ("fork holograms") that produce a set of diffracted beams with different OV charges. Usually, only single separate beams are used and investigated; here we consider the whole set of diffracted OV beams that, at certain conditions, are involved in efficient mutual interference to form a characteristic pattern where the ring-like structure of separate OV beams is replaced by series of bright and dark lines between adjacent diffraction orders. This pattern, well developed for high diffraction orders, reflects the main spatial properties of the diffracted beams as well as of the fork grating used for their generation. In particular, it confirms the theoretical model for the diffracted beams (Kummer beam model) and enables to determine the sign and the absolute value of the phase singularity embedded in the hologram.

Light beams with optical vortices (OV), or screw wavefront dislocations, have attracted a great attention during the past decades [1–5]. Their numerous research and practical applications have given rise to diverse methods for their generation, and a thin holographic grating with the groove bifurcation ("fork") is one of the simplest and the most universal practical means designed for this purpose [6–10]. Usually, when an incident paraxial beam with regular (non-vortex) wavefront (for example, a monochromatic Gaussian laser beam with the wavenumber $k$) intersects the central part of such a "fork grating" (FG) along the normal to its plane, a set ("fan") of diffracted beams with directions determined by angles

$$\theta_n = n\frac{2\pi}{kd} \tag{1}$$

depending on the diffraction order $n$ and the grating period $d$ is formed behind the grating (see Fig. 1). The $n$th-order diffracted beam carries the OV with topological charge

$$l_n = nq \tag{2}$$

where $q$ is the fixed topological charge of the phase singularity "embedded" in the grating. In this note we restrict ourselves to the case of integer $q$ where the "central" vertical groove is divided into ($|q|+1$) branches (in particular, in Fig. 1 $q=1$).

In general practice, only one of the diffracted beams with the desired OV charge is used; all other diffracted orders are filtered out and take no part in further manipulations. However, simultaneous observation of the multiple diffracted beams may provide some important information and gives a unique possibility to analyze the spatial structure of the separate diffracted beams as well as can serve to diagnostics of the properties of the generating diffraction element.

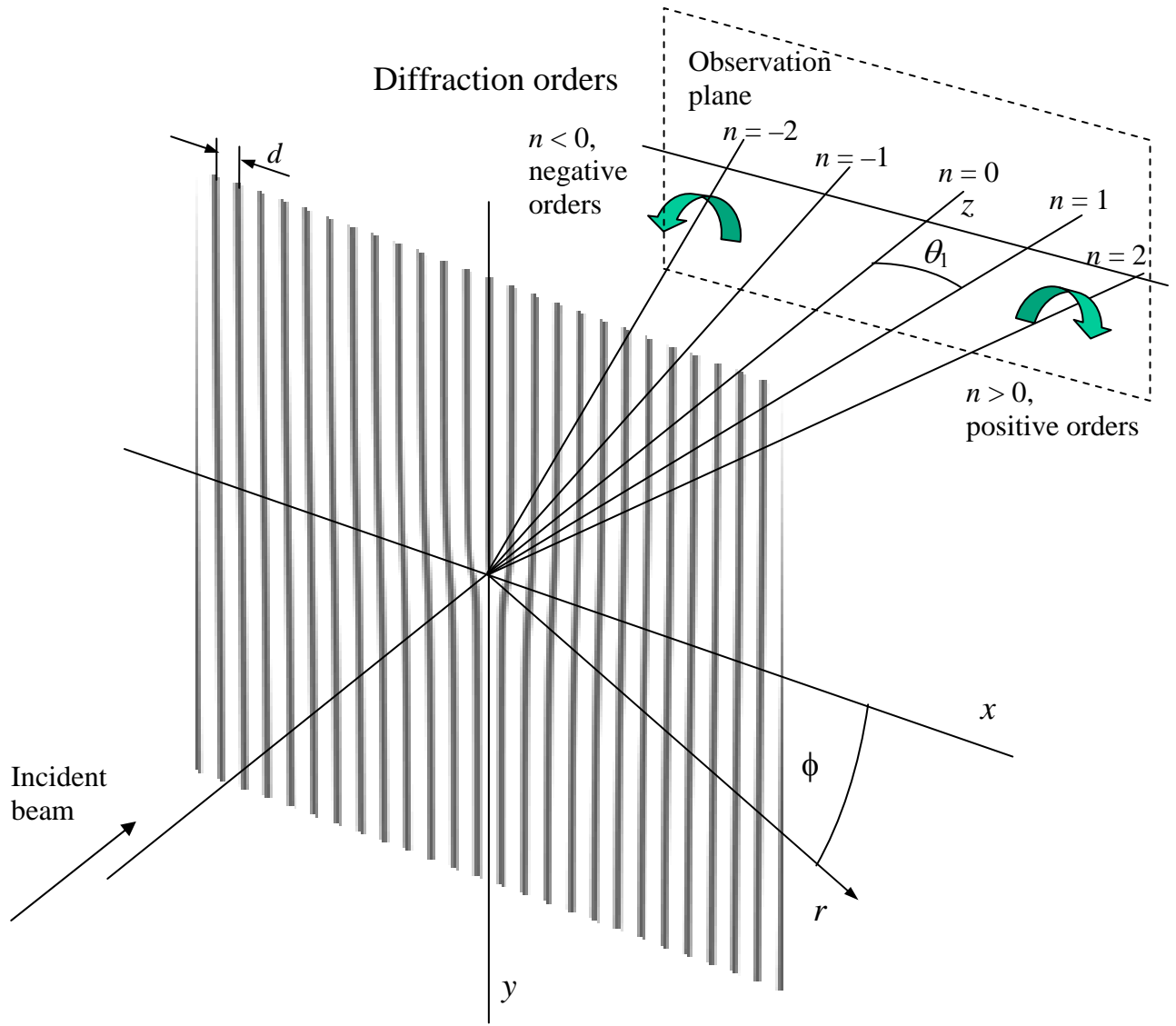

Fig. 1. Multiple OV beams produced by the FG. The Cartesian frame ($x$, $y$, $z$) is associated with the grating so that the axis $z$ is normal to the FG plane and intersects it exactly in the bifurcation point (the FG "center"), axis $y$ is parallel to the grooves far from the center. All axes of the diffracted beams belong to the diffraction plane ($XZ$).

Perhaps, every researcher that have ever dealt with a "simple" binary FG (without special groove profiling aimed at removal of some diffraction orders or al least at minimization of their available numbers), could observe the characteristic pattern of Fig. 2. Normally it cannot be seen in the vicinity of the incident beam axis where diffracted beams of the first few orders are concentrated, which are the most intense and most available for observation. But at the periphery of the diffracted "fan", the diffracted beams become to overlap, which manifests itself in the characteristic interference fringes "stretching" between the adjacent diffracted OV beams so that the expected ring-like beam spots are completely replaced by the quasiperiodic patterns of bright and dark strips (Fig. 2).

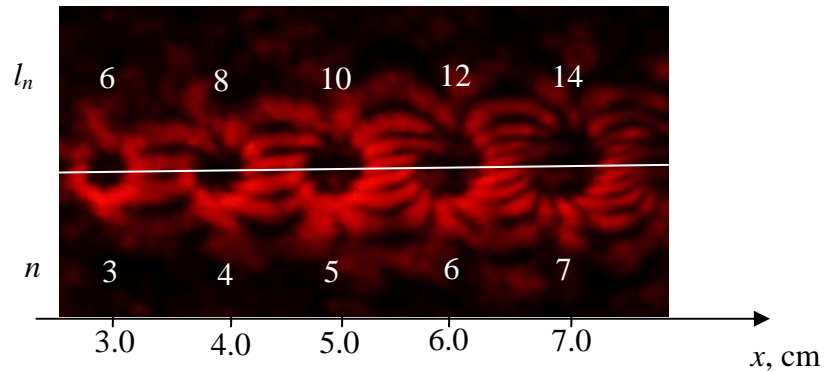

Fig. 2. The pattern of multiple OV beams formed after diffraction of a Gaussian laser beam with the waist radius $b_0 = 0.2$ mm in the FG with $q = 2$ in the $3^{rd} - 7^{th}$ diffracted orders observed on the screen distanced from the FG by $z = 100$ cm. Expected topological charges of the generated OV beams are indicated near the corresponding OVs, figures below indicate the diffraction orders; distance $x$ is measured from the zero-order beam center, white line is the trace of the diffraction plane. Other parameters of the experiment: $d = 1/16$ mm, $k \approx 10^5$ cm$^{-1}$ (He-Ne laser), $\theta_1 = 0.01$ rad.

Quite unexpectedly, we could find neither description nor interpretation of such pictures in the current literature. The main purpose of this note is to explain this pattern and to inspect how it can be used for investigation of the diffracted beams' spatial structure as well as for the FG structure diagnostics. We start with noticing that the spatial distribution of the complex amplitude of the $n$th-order diffracted beam at the screen plane can be expressed in the form

$$u_n(x, y, z) = U_n(r_n)\exp(il_n\varphi_n)\exp\left[i\frac{ka_n}{z}(x-a_n) + i\frac{k}{2z}a_n^2\right] \quad (3)$$

where $a_n = nz\theta_1$ is the position of the $n$th diffracted beam axis at the screen (the condition $\theta_1 \ll 1$ is implied),

$$r_n(x,y,z) = \sqrt{(x-a_n)^2 + y^2}, \quad \exp[i\varphi_n(x,y,z)] = \frac{(x-a_n) + i\sigma y}{r_n}, \quad (4)$$

$\sigma = \mathrm{sgn}(q) = \pm 1$ is the sign of the phase singularity embedded in the FG. The last exponential term in Eq. (3) expresses the additional phase of the slightly inclined $n$th-order beam in the observation plane normal to the nominal axis $z$ (see Fig. 1). The resulting observed field pattern can then be expressed as

$$I_{M,N}(x,y,z) = \left|\sum_{n=M}^{N} u_n(x,y,z)\right|^2 \quad (5)$$

where $M$ and $N$ are the minimum and maximum numbers of the diffraction orders whose influence cannot be neglected. Note that simultaneous replacement $x \to -x$, $n \to -n$ and $\sigma \to -\sigma$ does not change the value of expressions (3) and, consequently, (5) so that the resulting field pattern is symmetric with respect to the $y$-axis, which, in full agreement with the experimental observation, permits us to restrict the consideration by the positive $x$. For this reason, only positive diffraction orders, $n > 0$, will be considered below.

In most cases one may expect that in the region $a_n < x < a_{n+1}$ the field pattern can be evaluated via Eq. (5) by substitution $M = n$, $N = n+1$, i.e. allowance for only two adjacent diffracted beam is satisfactory. However, Eq. (5) provides a possibility to take into account the influence of not only of the nearest but also of long-distanced members of the diffracted beams' sequence.

Function $U_n(r_n)$ describes the details of the diffracted beam shape that depend on the model used for its characterization. In many cases, when the main attention is paid to the screw wavefront dislocation and the vortex properties of the FG-generated diffracted beam, it is suitable to describe it by the standard OV beam model supplied by the Laguerre-Gaussian (LG) mode [1,3,4]:

$$U_n(r_n) \equiv U_n^{LG}(r_n) = \left(\frac{r_n}{b}\right)^{|l_n|} \exp\left(-\frac{r_n^2}{2b^2}\right) \exp\left[ik\frac{r_n^2}{2R} - i(|l_n|+1)\chi\right] \quad (6)$$

where parameters of the beam radius $b$, wavefront curvature radius $R$ and the additional phase shift (Gouy phase) $\chi$ are conveniently fitted [11]. These are related with the "conventional" beam waist radius $b_1$ and corresponding conditional Rayleigh range $z_{R1} = kb_1^2$ that are chosen so that the formula (8) supply the best fitting to the real diffracted OV beam:

$$R(z) = \frac{z_{R1}^2 + z^2}{z}, \quad b^2(z) = \frac{z_{R1}^2 + z^2}{kz_{R1}}, \quad \chi(z) = \arctan\left(\frac{z}{z_{R1}}\right). \quad (7)$$

The LG approximation (6) can be useful for description of the spatial profile of a single diffracted beam, at least in separate cross sections [3,6,7,11,12]; however, it looses the validity in application to the "fan" of diffracted OV beams simultaneously produced by a single FG. Fig. 3 demonstrates that even in the situation where the beams efficiently overlap, the interference pattern calculated via the LG representation is quite different from the observed one: the fringes are almost rectilinear and strictly localized within the bright rings. In fact, the calculation of Fig. 3 was performed for $M = 3$, $N = 7$ in Eq. (5) but exactly the same picture could be obtained by combination of the interference patterns calculated for each pair of the nearest diffracted beams: this is a consequence of the very rapid exponential decay of the LG beam intensity with growing $r_n$.

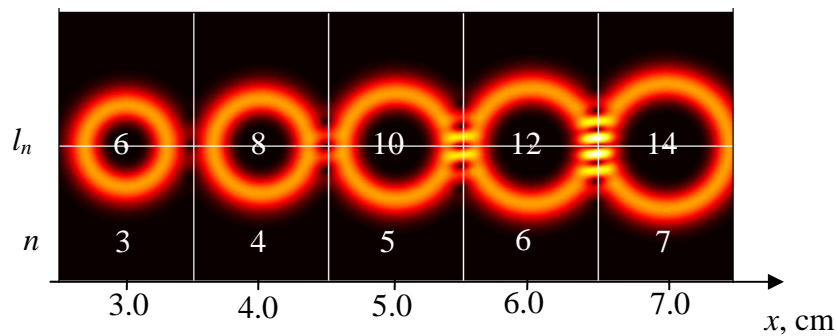

Fig. 3. The pattern of multiple OV beams formed by the FG with $q = 2$ in the 3$^{rd}$ to 7$^{th}$ diffracted orders in conditions corresponding to Fig. 2 calculated via Eq. (5) with $M = 3$, $N = 7$ for the LG model (6), (7). The conventional waist radius $b_1 = 0.08$ mm, vertical lines correspond to the middle distance between the diffracted beams' axes, horizontal line is the trace of the diffraction plane (see Fig. 1). Topological charges and corresponding diffraction orders are indicated as in Fig. 2, distance $x$ is measured from the zero-order beam center.

In this situation, one should resort to a more accurate model of the diffracted OV beam. As is known, for the case of a simple binary FG, the most adequate is the model of Kummer (hypergeometric-Gaussian) beam [10,11,13] where

$$U_n(r_n) \equiv U_n^K(r_n) = (-i)^{|l_n|+1} \exp\left(\frac{ik}{2z} r_n^2\right) \frac{z_R}{z - iz_R} \sqrt{A_n} \exp(-A_n) \left[ I_{\frac{|l_n|-1}{2}}(A_n) - I_{\frac{|l_n|+1}{2}}(A_n) \right], \quad (7)$$

$$z_R = kb_0^2 \quad (8)$$

is the Rayleigh range of the incident Gaussian beam whose waist ($b_0$ the waist beam radius at the level $e^{-1}$ of maximum intensity) is supposed to coincide with the FG plane, and

$$A_n = \left(\frac{kr_n}{z}\right)^2 \frac{b^2}{4(1 - iz_R/z)}. \quad (9)$$

The results of calculation by Eqs. (7) – (9) and (5) for $M = 3$, $N = 7$ are presented in Fig. 4, which should be compared with Fig. 2. Because the diffracted beams with $n < 3$ and $n > 7$ were not taken into account, Fig. 4 is expected to represent the correct results only in the region 3 cm $< x <$ 7 cm, including the 4th, 5th and 6th-order diffracted beams. Confronting this pattern with the corresponding area of Fig. 2 we conclude that the calculated picture is qualitatively similar to the experimental one, and the Kummer beam model of the diffracted beams really explains the observations.

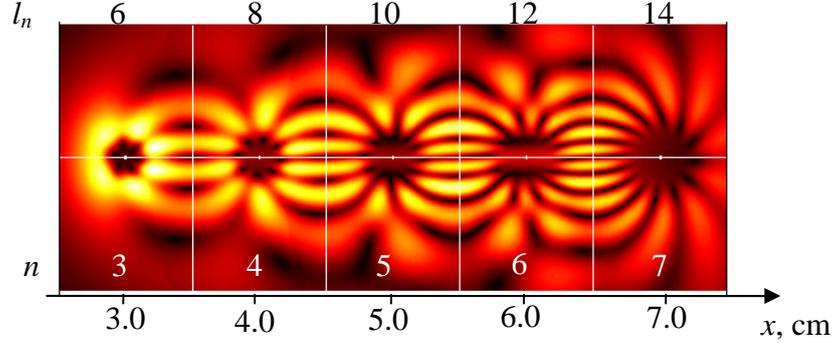

Fig. 4. The pattern of multiple OV beams formed by the FG with $q = 2$ in the 3rd to 7th diffracted orders in conditions corresponding to Fig. 2 calculated for the Kummer beam model (7) – (9) with $b_0 = 0.2$ mm. Topological charges, diffraction orders and distances $x$, $z$ are the same as in Figs. 2 and 3. The meaning of white lines is as in Fig. 3.

Expressions (3), (7) and (5) fairly describe the observed field structure but they are not suitable for analytical quantitative investigation. To make it more feasible, we may remark that the main details of the interference pattern (locations of dark and bright lines, etc.) are determined by the phase of the field (7) whereas its amplitude only specifies their relative brightness. Therefore, consider the "pure phase" part of Eq. (7) which can be treated as a version of general expression (3) with

$$U_n(r_n) \equiv U_n^P(r_n) = (-i)^{|l_n|+1} \exp\left(\frac{ik}{2z} r_n^2\right). \quad (10)$$

In this case one can easily find the "intensity distribution" taking into account only two nearest diffraction orders:

$$I_{n,n+1}(x,y,z) = 2\left\{1 + \cos\left[(l_{n+1}\varphi_{n+1} - l_n\varphi_n) - \frac{\pi}{2}(|l_{n+1}| - |l_n|)\right]\right\} \quad (11)$$

where Eqs. (4) are employed. Just at the middle distance between the axes of $n$th and $(n+1)$th diffracted beams with topological charges $l_n$ and $l_{n+1}$,

$$x = \frac{a_{n+1} + a_n}{2}, \tag{12}$$

the angular parameters $\varphi_n$ and $\varphi_{n+1}$ of Eq. (11) are determined by relations

$$\varphi_n \equiv \varphi_n^m = \arctan\frac{2y}{a_{n+1} - a_n}, \quad \varphi_{n+1} = \pi - \varphi_n^m. \tag{13}$$

In view of Eq. (2), $l_{n+1}\varphi_{n+1}^m - l_n\varphi_n^m = q(n+1)\pi - q(2n+1)\varphi_n^m$, and

$$I_{n,n+1} = 2 + 2\cos\left[(l_{n+1} + l_n)\varphi_n^m - \frac{\pi}{2}(|l_{n+1}| + |l_n|)\right] = 2 + 2\cos\left[(|l_{n+1}| + |l_n|)\left(\varphi_n^m - \frac{\pi}{2}\mathrm{sgn}(l_n)\right)\right]$$

$$= 2 + 2\cos\left[|q|(2n+1)\left(\varphi_n^m - \frac{\pi}{2}\mathrm{sgn}(q)\right)\right] = 2 + 2\begin{cases}(-1)^{q/2}\cos\left[|q|(2n+1)\varphi_n^m\right], & \text{if } q \text{ is even;} \\ (-1)^{n+(q-1)/2}\sin\left[|q|(2n+1)\varphi_n^m\right], & \text{if } q \text{ is odd.}\end{cases} \tag{14}$$

For known $n$, this gives the absolute value and sign of $q$. The absolute value $|q|$ can be determined from the number the intensity zeros or maximums situated along the vertical line (12). The number of zeros is

$$Z = \left\lfloor |q|\frac{2n+1}{2}\right\rfloor \tag{15}$$

where $\lfloor ... \rfloor$ is the symbol of the integer part of a number; the number of maximums is the same for odd $q$ and equals to $Z-1$ for even $q$.

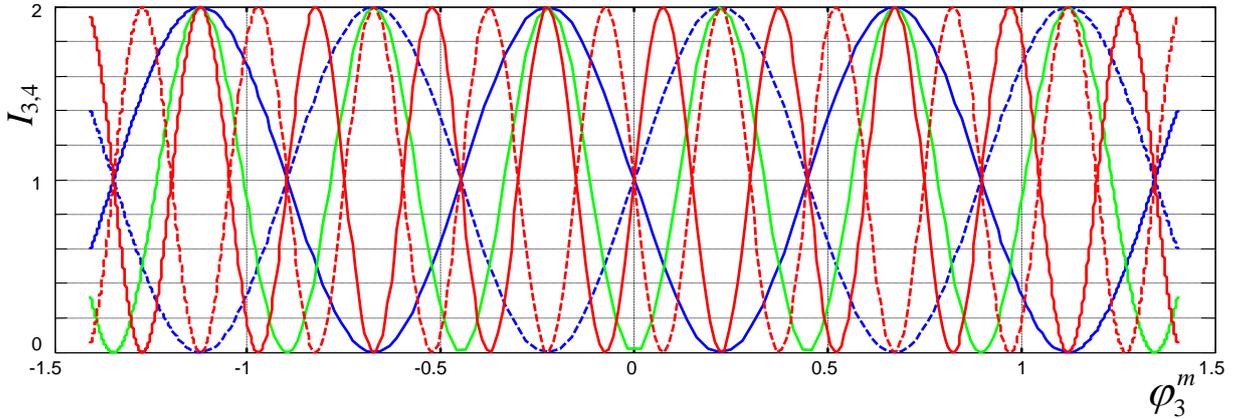

Fig. 5. Patterns of the intensity distribution (14) along the middle line (12) between the diffracted beams with orders $n = 3$ and $n = 4$. Solid lines: positive $q$, dashed lines: negative $q$; (blue) $|q| = 1$, (green) $|q| = 2$, (red) $|q| = 3$. The solid and dashed green lines coincide.

This situation is illustrated by Fig. 5. One may see that $|q|$ can be readily determined from the number of observed minima or maxima via Eq. (15); additionally, cases of positive and negative $q$ distinctly differ (the positions of maxima for $q > 0$ become positions of minima for $q < 0$, and vice versa). This does not work in case of even $q$: the solid and dashed green lines in Fig. 5 coincide and therefore the sign of even $q$ cannot be determined from the positions of intensity minima and maxima in the middle line (12). However, situations with positive and negative $q$ are rather different in other points between $x = a_n$ and $x = a_{n+1}$: comparison of Figs. 2, 3, 4 and 6 show that the sign of $q$ can be readily identified by the slight inclination of the interference fringes near the $x$-axis. Due to combined action of the screw and spherical components of the diffracted orders' wavefronts, the interference lines are not parallel to the $x$-axis: with growing $x$, these go slightly

'up' ('down') for positive (negative) $q$. Of course, this rule is also valid for odd $q$, and, again, can be used for the $q$ sign detection (which may appear more suitable in practice than tracing the alteration of minima and maxima along the middle line (12)).

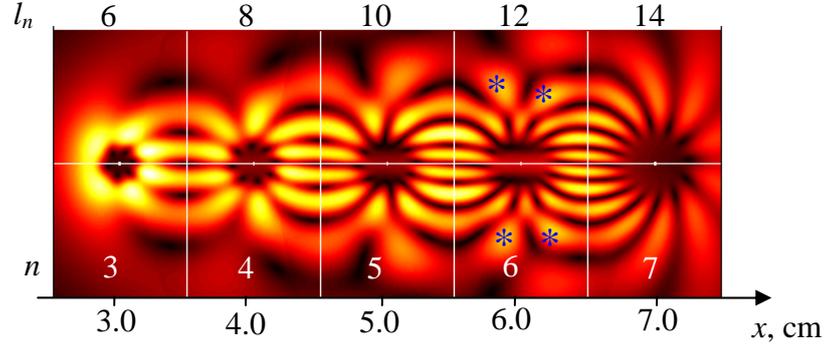

Fig. 6. The pattern of multiple OV beams formed by the FG with $q = -2$ in the 3rd to 7th diffracted orders; other conditions are the same as in Fig. 4.

Additional helpful rules can be seen in incremental growth of the interference fringes' number with increasing diffraction order. According to Eq. (15), the total number of zeros or maxima for consecutive orders $n$ and $n + 1$ grows by $|q|$. In conditions of Figs. 4 and 6 for lines $x = 3.5$ (between 3rd and 4th orders, $n = 3$), 4.5 ($n = 4$), 5.5 ($n = 5$),… the number of zeros amounts to 7, 9, 11,…, correspondingly, that is, increment is $2 = |q|$. Likewise, in Fig. 7 this increment expectedly equals 1, and in Fig. 8 it is 3.

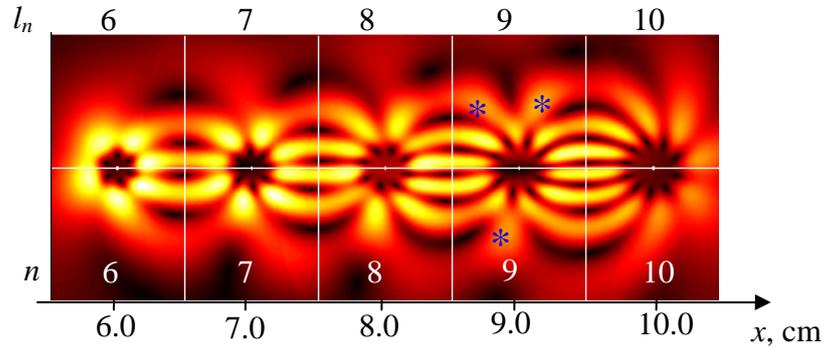

Fig. 7. The pattern of multiple OV beams formed by the FG with $q = 1$ in the 6th to 10th diffracted orders calculated for the Kummer beam model (7) – (9) with $b_0 = 0.2$ mm. The distance $z$ is the same as in Figs. 2 – 4 and 6, distance $x$ is measured from the zero-order beam axis. The meaning of white lines is as in Fig. 3.

The presented regularities can be used, e.g., for the express diagnostics of FGs, i.e. if one needs to learn the topological charge $q$ of the embedded phase singularity. However, the detection of some intensity minima or maxima far from the $x$-axis can be practically difficult because of lowering the diffracted beam intensity with growing $y$. For example, in Figs. 4 and 6 ($q = 2$), the number of zeros along vertical lines $x = 3.5$ cm ($n = 3$) and $x = 4.5$ cm ($n = 4$) dictated by Eq. (15) should be 7 and 9 while only 5 and 7 minima are really seen: The intensity minima situated at high $|y|$ are hardly discernible in practice because of the low local light intensity and inevitable noise contamination. In such cases, observation of interference pattern within a limited vertical segment can be purposeful. If one takes only the brightest interference lines into account, in Fig. 2 their numbers are 3, 5, 7 (between 3rd and 4th, 4th and 5th, 5th and 6th orders); in Figs. 4 and 6 one sees 4, 6, 8 lines, which clearly testifies for $|q| = 2$. This criterion seems to be the most stable against practically inevitable noise and distortion: in practical experiments, due to the presence of many orders, it is usually not difficult to find several of them with which the incremental growth of the fringe numbers can be reliably identified.

In more accurate measurements, positions of the bright lines near the *x*-axis can also be helpful. In accordance to the last expression (14), the regularities are different for even and odd $|q|$. In case of even $|q|$, for every *n* the interference pattern possesses an extremum at $y = 0$: minimum for $|q| = 2, 6, 10, \ldots$ and maximum for $|q| = 4, 8, 12, \ldots$ (cf. Figs. 1, 4 and 6). In case of odd $|q|$, the maxima and minima are situated just above or just below the horizontal axis $y = 0$, and, for $|q| = 1, 5, 9, \ldots$, the intensity above the *x*-axis is maximum at even *n* and minimum at odd *n* (cf. Fig. 7, the interference lines between $6^{th}$ and $7^{th}$, $7^{th}$ and $8^{th}$, $8^{th}$ and $9^{th}$, $9^{th}$ and $10^{th}$-order diffracted beams). For $|q| = 3, 7, 11, \ldots$, the vertical intensity distribution along the middle line (12) shows an inverse behaviour: above the *x*-axis one can see minima at even *n* and maxima at odd *n* (cf. Fig. 8, interference lines between $2^{nd}$ and $3^{rd}$, $3^{rd}$ and $4^{th}$, $4^{th}$ and $5^{th}$, $5^{th}$ and $6^{th}$-order diffracted beams).

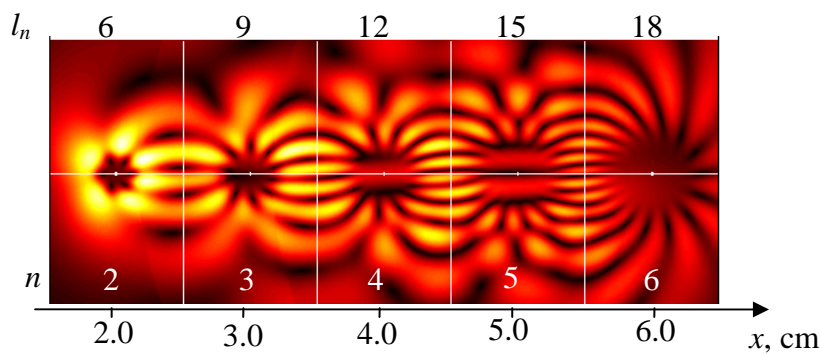

Fig. 8. The pattern of multiple OV beams formed by the FG with $q = 3$ in the $2^{nd}$ to $6^{th}$ diffraction orders calculated for the Kummer beam model (7) – (9) with $b_0 = 0.2$ mm. The distance *z* is the same as in Fig. 7, distance *x* is measured from the zero-order beam axis. The meaning of white lines is as in Fig. 3.

Other possible approaches to the FG diagnostics can employ the apparent fact visible in all figures that, due to superposition of beams of different diffraction orders, an expected ring-like structure in each order is "fractioned" into a "necklace" of bright spots whose number systematically grows with *n* and $|q|$. However, the considered examples show no distinct quantitative regularity in this growth: while in Figs. 4 and 6 the rings corresponding to $n = 4$, 5 and 6 are split into 8, 10 and 12 spots, respectively, in Fig. 7 the rings of $n = 8$ and 9 are both split into 10 spots. In the situation of Fig. 8 the rings of orders $n = 3$, 4 and 5 are decomposed into 10, 14 and 18 spots (a sort of regularity seems to be restored but the increment value 4 is not understandable). Besides, the experimental identification of the separate spots within the presumed rings is sometimes difficult. E.g., in Figs. 6 and 7 there are three bright spots within the expected $6^{th}$ and $9^{th}$-order "rings", marked by asterisks. They are positioned at higher distances from the beam axis than other ones, for which reason they were not taken into account when reckoning the number of the ring "fragments" a few lines above. However, in real experiments, inevitable distortions of the theoretical pattern can make these maxima visually equivalent to the others so that calculation of the bright spots may be ambiguous. Probably, for this reason the numbers of bright spots in "necklaces" seen in Fig. 2 differ from the theoretical predictions of Figs. 4 and 6.